\documentclass[11pt,a4paper]{article}

\usepackage{amssymb}
\usepackage{amsmath}
\usepackage[pdftex]{graphicx}
\usepackage{graphics}
\usepackage{epsfig}
\usepackage{color}
     \usepackage[colorlinks]{hyperref}
\hypersetup{linkcolor=blue,%
citecolor=red,%
urlcolor=cyan}
\usepackage{url} 
 
\usepackage[parfill]{parskip}    

\newcommand{\be}{\begin{equation}}
\newcommand{\ee}{\end{equation}}
\numberwithin{equation}{section}

\newtheorem{rem}{Remark}

\title{\bf The Schr\"odinger particle on the half-line with an attractive $\delta$-interaction: bound states and resonances}

\author{S. Fassari$^{1,2,3}$\footnote{silvestro.fassari@uva.es, ORCID: \href{http://orcid.org/0000-0003-3475-7696}{0000-0003-3475-7696}}, 
M. Gadella$^4$\footnote{manuelgadella1@gmail.com, ORCID: \href{http://orcid.org/0000-0001-8860-990X}{0000-0001-8860-990X}},  
L.M. Nieto$^4$\footnote{luismiguel.nieto.calzada@uva.es, ORCID: \href{http://orcid.org/0000-0002-2849-2647}{0000-0002-2849-2647}}, 
 and 
F. Rinaldi$^{2,3}$\footnote{f.rinaldi@unimarconi.it, ORCID:  \href{http://orcid.org/0000-0002-0087-3042}{0000-0002-0087-3042}}
\\ [2ex]
\footnotesize \sl $^1$Department of Higher Mathematics, ITMO University, S. Petersburg, Russian Federation\\
\footnotesize \sl $^2$Univ. degli Studi Guglielmo Marconi,Via Plinio 44, I-00193 Rome, Italy \\
\footnotesize \sl $^3$CERFIM, PO Box 1132, CH-6601 Locarno, Switzerland \\
\footnotesize \sl $^4$Departamento de F\'{\i}sica Te\'{o}rica, At\'{o}mica y \'{O}ptica, and IMUVA,\\ 
\footnotesize \sl U. de Valladolid, 47011 Valladolid, Spain.
}

\begin{document}

\maketitle

\begin{abstract}
 In this paper we provide a detailed description of the eigenvalue $ E_{D}(x_0)\leq 0$ (respectively $ E_{N}(x_0)\leq 0$) of the self-adjoint Hamiltonian operator representing the negative Laplacian on the positive half-line with a Dirichlet (resp. Neuman) boundary condition at the origin perturbed by an attractive Dirac distribution $-\lambda \delta(x-x_0)$ for any fixed value of the magnitude of the coupling constant. We also investigate the $\lambda$-dependence of both eigenvalues for any fixed value of $x_0$. Furthermore, we show that both systems exhibit resonances as poles of the analytic continuation of the resolvent.  These results will be connected with the study of the ground state energy of two remarkable three-dimensional self-adjoint operators, studied in depth in Albeverio's monograph, perturbed by an attractive $\delta$-distribution supported on the spherical shell of radius $r_0$.
\end{abstract}

\bigskip

\section{Introduction}
Point interactions have drawn an increasing amount of interest in Theoretical and Mathematical Physics over the last six decades, mainly because they can be used in place of sharply peaked potentials to obtain solvable models in Quantum Mechanics (see \cite{AGHH} and a more thoroughly mathematical discussion in \cite{AK}). As a consequence, they provide a wide range of solvable models in Quantum Mechanics, which may serve as a laboratory to investigate quantum properties of a variety of quantum systems including unstable quantum systems. The present article, dealing with a model that shows both bound states and resonances, attempts to provide a new contribution to the subject.

In relation with one-dimensional point interactions, we must say that there is an extensive  literature concerning both mathematical and physical properties of such potentials \cite{AGHH,AK,ADK,KUR,ZOL,ZOL1,ZOL2,GOL,EU,LUN,LUN1,SF,BR}. From the mathematical point of view,  such one-dimensional singular interactions/potentials may be associated with self-adjoint extensions of the differential operator $-d^2/dx^2$ on a given domain where this operator is symmetric with equal non-zero deficiency indices. The most popular of these interactions is the Dirac delta $\delta(x-x_0)$, which may appear alone (see the above-mentioned works), in the form of a system of a finite or infinite number of deltas forming a crystal \cite{AGHH,EGU}, with a mass jump or as a point perturbation of a Hamiltonian of the form $-d^2/dx^2+V(x)$.  A perturbation of the type $\delta'(x-x_0)$ exists in the current literature \cite{ZOL3,ZOL4} as defined by two distinct self-adjoint extensions of $-d^2/dx^2$. One gives the non-local $\delta'$- interaction \cite{ENZ,PSEB,AFR3,AFR4,FGGN} and the other is a local interaction compatible with the delta at the same point, so that we may construct interactions of the type $\alpha\,\delta(x-x_0) + \beta \, \delta'(x-x_0)$ for $-d^2/dx^2$ (or even for $-d^2/dx^2+V(x)$, where in the simplest cases $V(x)$ is a harmonic or conic oscillator potential, or an infinite square well  \cite{FGGN,GGN,GGN1}). In addition, there are other types of contact potentials, for which their physical meaning is under discussion \cite{KP,KP1}. Delta perturbations may also be added to relativistic or semirelativistic free particle Hamiltonians, although in this case a regularisation process is essential in order to have a perfectly defined self-adjoint Hamiltonian \cite{AK1,HSW,AFR,EGU1}. 

Concerning physical applications of point potential interactions, these are multiple. Let us mention: a Bose-Einstein condensation in a harmonic trap with a tight and deep dimple potential, modelled by a Dirac delta \cite{UTDM}, a nonperturbative study of the entanglement of two directed polymers subjected to repulsive interactions given
by a Dirac delta \cite{FRV}, a study on light propagation in a one-dimensional realistic dielectric superlattice, modelled by means of a periodic array of Dirac delta functions \cite{ALV,ZUR,LIN},  the implementation of quantum dots \cite{HAUS,HJO,KVA}, or plasma frequencies in the Barton hydrodynamical model \cite{BAR}.  The applications in Quantum Field Theory are also numerous \cite{MUN,ASO,ASO1,MUN1,MAT,BOR}, for example Casimir physics has also been interpreted on the light of contact interactions \cite{BOR1,FK,FU}. Finally, the merging of contact interactions may also have a group theoretical interpretation \cite{GMMN}.

As is well known, three-dimensional systems with spherical symmetry can be reformulated as one-dimensional systems with support along the positive real half-line. The Laplacian Hamiltonian perturbed by a delta distribution supported on a spherical shell has been investigated in the literature \cite{AGS,SHA,DS,CNR}. The main objective of the present paper is to provide a model of the latter in one dimension (zero orbital angular momentum) in two different versions characterised by two remarkable boundary conditions  at the origin: Dirichlet and Neumann. Some precedents of our investigation can be found in \cite{DER,SEB,ES}.

Thus, we show that the operator $-d^2/dx^2$ on $\mathbb R^+\equiv [0,\infty)$ with an attractive Dirac delta perturbation  situated at $x_0>0$  plus a Dirichlet or Neumann boundary condition at the origin is a rigorously defined 
self-adjoint operator,  endowed with a bound state and an infinite number of resonances. Although there are several definitions of resonances, not all equivalent \cite{BOHM,FGR,NUSS}, we are using here the following~\cite{RSIV}:
\begin{itemize}
\item[]
Let $\{H_0,H=H_0+V\}$ be a Hamiltonian pair. Assume that there is a dense subspace $\mathcal D$ such that for each $\psi\in\mathcal D$, the functions
\begin{equation}\label{1.1}
R^0_\psi(E)=\Big\langle \psi \Big|  \frac{1}{H_0-E}\,\psi \Big\rangle\,, \qquad 
R_\psi(E)=\Big\langle \psi \Big|  \frac{1}{H-E}\,\psi \Big\rangle\,,
\end{equation}
admit analytic continuation for $E$ complex. If for some $\psi\in\mathcal D$, $R^0_\psi(E)$ is analytic at $z_0=E_0-i\Gamma/2$, $\Gamma>0$, and $R_\psi(E)$ shows a pole at $z_0$, then, we say that the Hamiltonian pair $\{H_0,H=H_0+V\}$ exhibits a resonance at $z_0$. 
\end{itemize}

This paper is organised as follows: in Sections 2 and 3 we construct the resolvent $(H-E)^{-1}$ of the Hamiltonian $H=-d^2/dx^2-\lambda\delta(x-x_0)$, $x_0,\lambda>0$, on $L^2(\mathbb R^+)$ with Dirichlet and Neumann boundary conditions at the origin, respectively. We give the conditions for the emergence of  an eigenvalue below the continuous spectrum and give its energy as a function of $x_0$ (espectively $\lambda$) for different values of the coupling constant $\lambda$ (respectively $x_0$). We also show the existence of resonances. In Section 4, we include some comments on the three-dimensional Hamiltonian $-\Delta-\lambda \delta(r-r_0)$, where $r$ is the radial coordinate and $r_0>0$. We finish this article with some concluding remarks.

\section{The operator $\left[-\frac {d^2}{dx^2}\right]_{D^+}$ perturbed by an attractive Dirac delta}
As is well known (see \cite{DER}), the self-adjoint Hamiltonian $\left[-\frac {d^2}{dx^2}\right]_{D^+}$, that is to say the Dirichlet Laplacian on the positive half-line with a Dirichlet boundary condition at the origin, has its resolvent given by the integral operator whose kernel, known in the more physically oriented literature as Green's function, is for any $x,y>0$ and $E<0$:
\begin{equation}\label{2.1}
\left[\left[-\frac {d^2}{dx^2}\right]_{D^+}\!\!+ |E|\right]^{-1}\!\!(x,y)=\frac {e^{- |E|^{1/2} |x-y|}-e^{- |E|^{1/2}(x+y)}}{2 |E|^{1/2}}\,.
\end{equation}
It is worth mentioning that this integral kernel, extended to the entire real line by having $x,y$ replaced by their absolute values in the exponent of the second term in the numerator, plays a crucial role in the study of the Birman-Schwinger operator for one-dimensional potentials whose $(1+\epsilon)$-moment is integrable (see, e.g., \cite{KlausHPA,Klaus,F2}). Furthermore, the integral kernel of the semigroup is also explicitly given in \cite{DER} (see \cite{Rend,ROMP15} as well):
\begin{equation}\label{2.2}
e^{-t\left[-\frac {d^2}{dx^2}\right]_{D^+}}(x,y)=\frac {e^{- \frac {(x-y)^2}{4t}}-e^{- \frac {(x+y)^2}{4t}}}{2 \sqrt{\pi t}}\,.
\end{equation}

We now wish to consider what happens when the operator gets perturbed by an attractive Dirac distribution centred at $x_0>0$ with strength $\lambda>0$, so that the new Hamiltonian reads:
\begin{equation}\label{2.3}
H_{\lambda,x_0}^{D}=\left[-\frac {d^2}{dx^2}\right]_{D^+}\!\!-\lambda \delta (x-x_0)\,.
\end{equation}

Differently from other related publications \cite{DER,SEB,ES} considering a delta distribution or a potential with shrinking support centred at the origin only, here the $\delta$-distribution is supported at the point $x_0>0$. 

Our next goal is to obtain an explicit expression for the resolvent of the perturbed Hamiltonian, that is
\begin{equation}\label{2.4}
\left[\left[-\frac {d^2}{dx^2}\right]_{D^+}-\!\!\lambda \delta (x-x_0)+ |E|\right]^{-1}\,
\end{equation}
We start by noticing that
\begin{eqnarray}\label{2.5}
\lambda \left[\left[\left[-\frac {d^2}{dx^2}\right]_{D^+}\!\!+ |E|\right]^{-1/2}\delta (x-x_0)\left[\left[-\frac {d^2}{dx^2}\right]_{D^+}\!\!+ |E|\right]^{-1/2}\right](x,y)=\nonumber \qquad\qquad\\
=\lambda \left[\left[-\frac {d^2}{dx^2}\right]_{D^+}\!\!+ |E|\right]^{-1/2}\!\! (x,x_0)\left[\left[-\frac {d^2}{dx^2}\right]_{D^+}\!\!+ |E|\right]^{-1/2}\!\!(x_0,y)\,,
\end{eqnarray}
which obviously defines the integral kernel of a rank one operator. The integral kernel of the square of \eqref{2.5} is given by:
\begin{equation*}\label{2.6}
\left[\left[-\frac {d^2}{dx^2}\right]_{D^+}\!\!+ |E|\right]^{-1}\!\!(x_0,x_0)\left[\left[-\frac {d^2}{dx^2}\right]_{D^+}\!\! + |E|\right]^{-1/2}\!\!(x,x_0)\left[\left[-\frac {d^2}{dx^2}\right]_{D^+}\!\!+ |E|\right]^{-1/2}\!\!(x_0,y)\,,
\end{equation*}
which is the kernel of the rank one operator in \eqref{2.5} times the integral kernel of the resolvent of $\left[-\frac {d^2}{dx^2}\right]_{D^+}$ evaluated at $x=y=x_0$. Hence, the kernel of the $\ell-$th power of \eqref{2.4} reads:
\begin{equation*}\label{2.7}
\left[ \left[\left[-\frac {d^2}{dx^2}\right]_{D^+}\!\!+ |E|\right]^{\!\!-1}\!\!\!(x_0,x_0)\right]^{l-1} 
\left[\left[-\frac {d^2}{dx^2}\right]_{D^+}\!\!+ |E|\right]^{\!\!-1/2}\!\!\!(x,x_0)\left[\left[-\frac {d^2}{dx^2}\right]_{D^+}\!\!+ |E|\right]^{\!\!-1/2}\!\!\!(x_0,y)\,.
\end{equation*}

Thus, it is now straightforward to compute the integral kernel of the Neumann expansion for \eqref{2.4}, which is given by
\begin{eqnarray}
\sum_{\ell=1}^{\infty} \lambda^\ell \left[ \left[\left[-\frac {d^2}{dx^2}\right]_{D^+}\!\!+ |E|\right]^{\!-1}\!\!\!(x_0,x_0)\right]^{\ell-1} 
\left[ \left[-\frac {d^2}{dx^2}\right]_{D^+}\!\!+ |E|\right]^{\!-1/2} \!\!\!(x,x_0)   \left[\left[-\frac {d^2}{dx^2}\right]_{D^+}\!\!+ |E|\right]^{\!-1/2} \!\!\!(x_0,y)  \nonumber \\
=\lambda \sum_{\ell=0}^{\infty} \left[ \lambda \left[\left[-\frac {d^2}{dx^2}\right]_{D^+}\!\!+ |E|\right]^{\!-1}\!\!\!(x_0,x_0)\right]^{\ell}   \left[ \left[-\frac {d^2}{dx^2}\right]_{D^+}\!\!+ |E|\right]^{\!-1/2} \!\!\!(x,x_0)  
 \left[\left[-\frac {d^2}{dx^2}\right]_{D^+}\!\!+ |E|\right]^{\!-1/2}\!\!\! (x_0,y)  \nonumber \\
 =\frac {\lambda}{1-\lambda \left[\left[-\frac {d^2}{dx^2}\right]_{D^+}\!\!+ |E|\right]^{\!-1}\!\!\! (x_0,x_0)} 
\left[ \left[-\frac {d^2}{dx^2}\right]_{D^+}\!\!+ |E|\right]^{\!-1/2}\!\!\! (x,x_0)  \left[\left[-\frac {d^2}{dx^2}\right]_{D^+}\!\!+ |E|\right]^{\!-1/2} \!\!\!(x_0,y).
\label{2.8} 
\end{eqnarray}

In order to obtain the resolvent \eqref{2.4} of the total Hamiltonian $H^D_{\lambda,x_0}$ \eqref{2.3}, we have to multiply \eqref{2.8} to the left and right by $\left[\left[-\frac {d^2}{dx^2}\right]_{D^+}\!\!+ |E|\right]^{-1/2}$, so that for $E<0$ the integral kernel  of the resolvent is given by:
\begin{eqnarray} \label{2.9}
(H^D_{\lambda,x_0}-E)^{-1}(x,y)= \left[\left[-\frac {d^2}{dx^2}\right]_{D^+}\!\!-\lambda \delta (x-x_0)+ |E|\right]^{-1}\!\!(x,y)=  \left[\left[-\frac {d^2}{dx^2}\right]_{D^+}\!\!+ |E|\right]^{-1} \!\!(x,y) \qquad \nonumber \\
+ \frac { \lambda}{1-\lambda \left[\left[-\frac {d^2}{dx^2}\right]_{D^+}\!\!+ |E|\right]^{-1}\!\!(x_0,x_0)} 
  \left[ \left[-\frac {d^2}{dx^2}\right]_{D^+}\!\!+ |E|\right]^{-1} \!\!(x,x_0) 
\left[\left[-\frac {d^2}{dx^2}\right]_{D^+}\!\!+ |E|\right]^{-1} \!\!(x_0,y) \,.
\end{eqnarray}
Needless to say, \eqref{2.9} is in full agreement with the well-known Krein formula (see \cite{AGHH,AK}). Therefore, since the eigenvalues of the operator are given by real poles of its resolvent, the equation determining the single eigenvalue of $H^D_{\lambda,x_0}$ is:
\begin{equation}\label{2.10}
\lambda \left[\left[-\frac {d^2}{dx^2}\right]_{D^+}\!\!+ |E|\right]^{-1}\!(x_0,x_0)=\lambda\ \frac {1-e^{- 2|E|^{1/2}x_0}}{2 |E|^{1/2}}=1\,.
\end{equation}

It is noteworthy that the last relation in \eqref{2.10} is precisely equation (2.10) in \cite{FR1}, which is the equation determining the energy of the excited bound state of the self-adjoint Hamiltonian
\begin{equation}\label{2.11}
-\frac {d^2}{dx^2}-\lambda [\delta (x+x_0)+\delta (x-x_0)]\,. 
\end{equation}
This Hamiltonian \eqref{2.11} is defined on a dense subset of the space of square integrable functions \textit{on the entire real line}, $L^2(\mathbb R)$. The crucial point of the analysis carried out in that article is that the emergence of this eigenvalue out of the absolutely continuous spectrum takes place provided $\lambda x_0>1$. As we wish to obtain the eigenenergy as a function of $x_0$ for any fixed value of $\lambda$, we note that \eqref{2.10} can be rewritten as
\begin{equation}\label{2.12}
x_0^{D}(E)=-\frac 1{2 |E|^{1/2}} \text{ln} \left(1- \frac {2 |E|^{1/2}}{\lambda}\right)\,.
\end{equation}

\begin{figure}[tb]
\begin{center}
\includegraphics[width=0.34\textwidth]{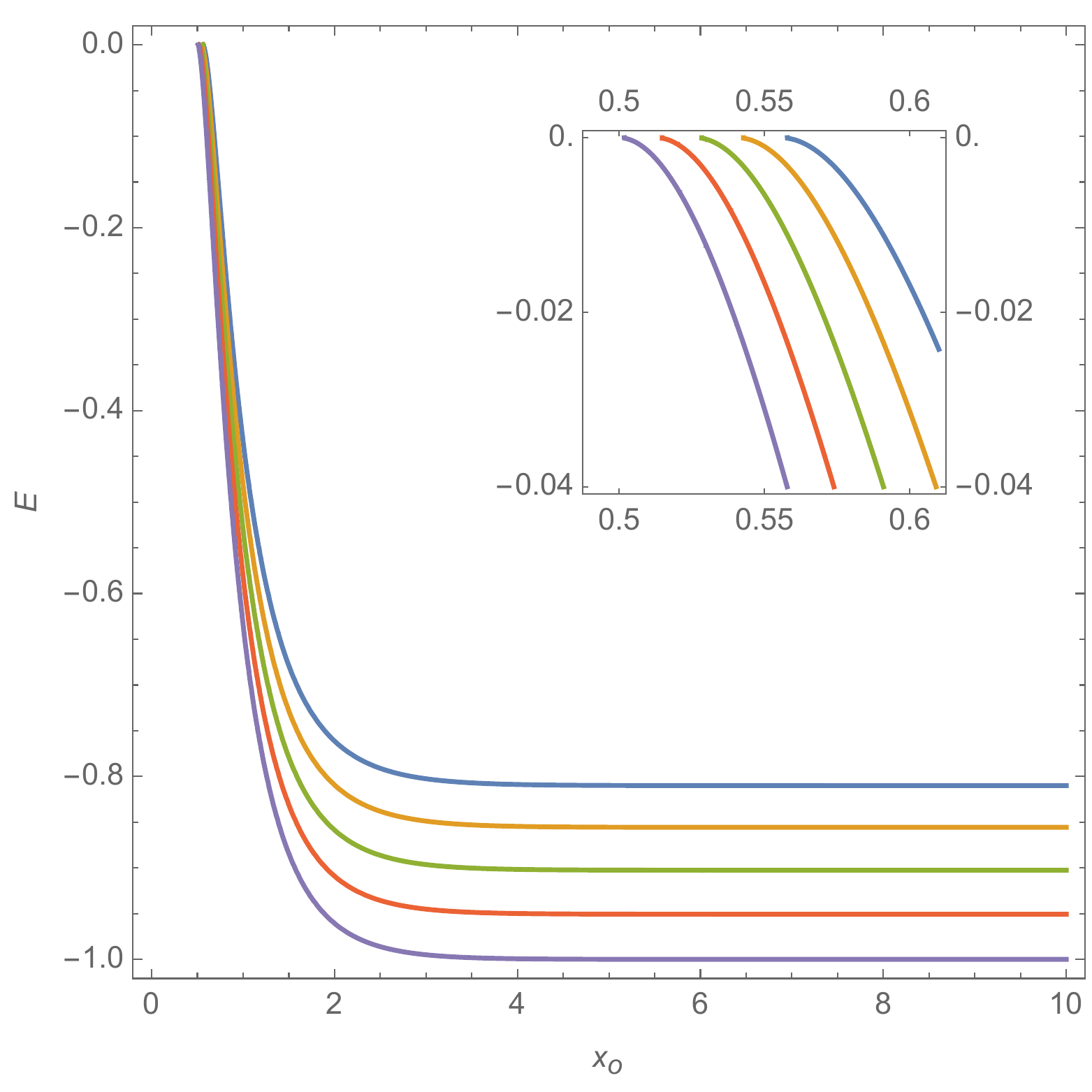}
\qquad\qquad\qquad
\includegraphics[width=0.34\textwidth]{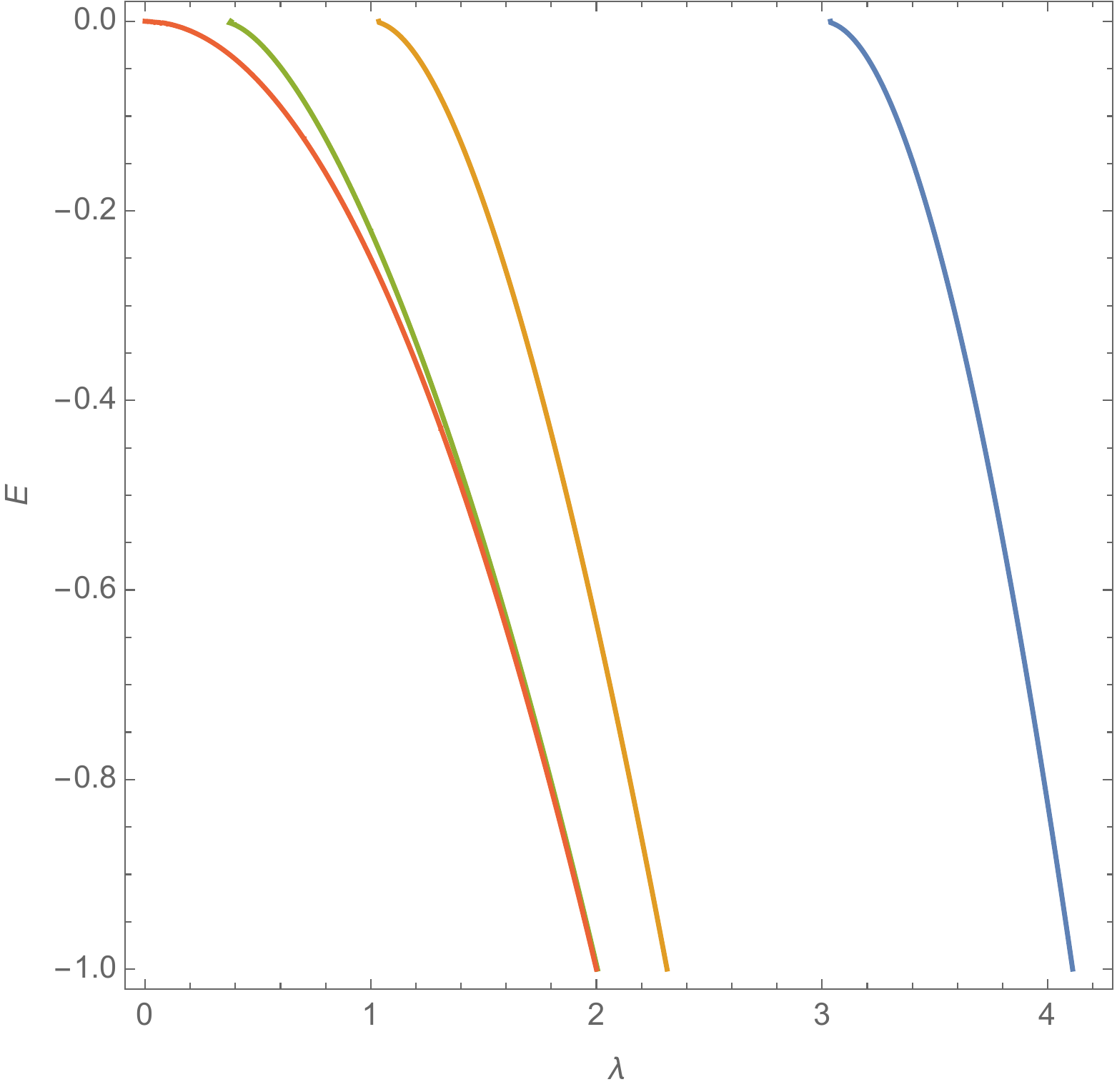}
\caption{On the left, the  energy $E_{D}(x_0)$ resulting from \eqref{2.10} for  
$\lambda=1.8$ (blue curve), $\lambda=1.85$ (orange curve), $\lambda=1.90$ (green curve), $\lambda=1.95$ (red curve), and $\lambda=2$ (violet curve). The inset shows the behavior of the curves near $E=0$.
On the right, the  energy $E_{D}(\lambda)$ from the solution of \eqref{2.10} for  
$x_0=1/3$ (blue curve), $x_0=1$ (orange curve), $x_0=3$ (green curve), and $x_0=\infty$ (red curve).}
\label{Dirichlet2D}
\end{center}
\end{figure}

For any fixed value of $\lambda$, the function $x_0^{D}(E)$ exists only for $-\frac {\lambda^2}{4}<E\leq 0$. Then, whilst $x_0(E) \rightarrow +\infty$ as $E \rightarrow -\frac {\lambda^2}{4}$ from above, $x_0^{D}(E) \rightarrow \frac 1{\lambda}$ as $E \rightarrow 0$ from below, as can be easily gathered by using the remarkable limit $ \text{lim}_{x \rightarrow 0} (1+x)^{1/x}=e$. Hence, given that $x_0^{D}(E)$ is a strictly decreasing function over its domain $\left(-\frac {\lambda^2}{4}, 0\right]$, its inverse $E_{D}(x^D_0)$ is well defined over the semibounded interval $ \left[\frac 1{\lambda}, +\infty \right)$ with range $ \left(-\frac {\lambda^2}{4},0 \right]$. Therefore, assuming $\lambda$ to be fixed, the eigenvalue $E_{D}(x_0)\leq 0$ exists only for $ x_0\geq \frac 1{\lambda}$ and $E_{D}(\frac 1{\lambda})= 0$. On the left hand side of Figure~\ref{Dirichlet2D} we plot this eigenvalue $E_{D}(x_0)$ as a function of the distance $x_0$ for several values of $\lambda$.

Now, take
\begin{equation}\label{2.13}
\left[\left[-\frac {d^2}{dx^2}\right]_{D^+}\!\!+ |E|\right]^{-1}\!\!(\!x_0,x_0) \quad \text{and}
\quad 
\left[\left[-\frac {d^2}{dx^2}\right]_{D^+}\!\!+ |E|\right]^{-1}\!\!(x,x_0) \left[\left[-\frac {d^2}{dx^2}\right]_{D^+}\!\! + |E|\right]^{-1}\!\!(x_0,y)\,.
\end{equation}
Both expressions in \eqref{2.13} approach 0 as $x_0 \rightarrow 0_{+}$. Then, it is straightforward to check that, for any fixed $\lambda$, the rank one operator on the right hand side of \eqref{2.9} vanishes, a fact which implies the norm resolvent convergence, defined in \cite{RSI}, of $H_{\lambda,x_0}^{D}=\left[-\frac {d^2}{dx^2}\right]_{D^+}\!\!-\lambda \delta (x-x_0)$ to the unperturbed Hamiltonian $\left[-\frac {d^2}{dx^2}\right]_{D^+}$ as $x_0 \rightarrow 0_{+}$, the spectrum of which is purely absolutely continuous. Here, we have performed an $x_0$-limit as in \cite{SF,ENZ,AFR4,AFR,FR1,AFR1,AFR2,FPR}.

The limit as $x_0 \rightarrow +\infty$ can also be obtained taking into account that
\begin{equation}\label{2.14}
\lambda \left[\left[-\frac {d^2}{dx^2}\right]_{D^+}\!\!+ |E|\right]^{-1}\! (x_0,x_0)=\lambda\ \frac {1-e^{- 2|E|^{1/2}x_0}}{2 |E|^{1/2}} \rightarrow \lambda\ \frac {1}{2 |E|^{1/2}}\,,
\end{equation}
so that $E_{D}(x_0)$ approaches $-\frac {\lambda^2}{4}$ for large values of $x_0$.

\begin{figure}[hbtp]
\begin{center}
\includegraphics[width=0.34\textwidth]{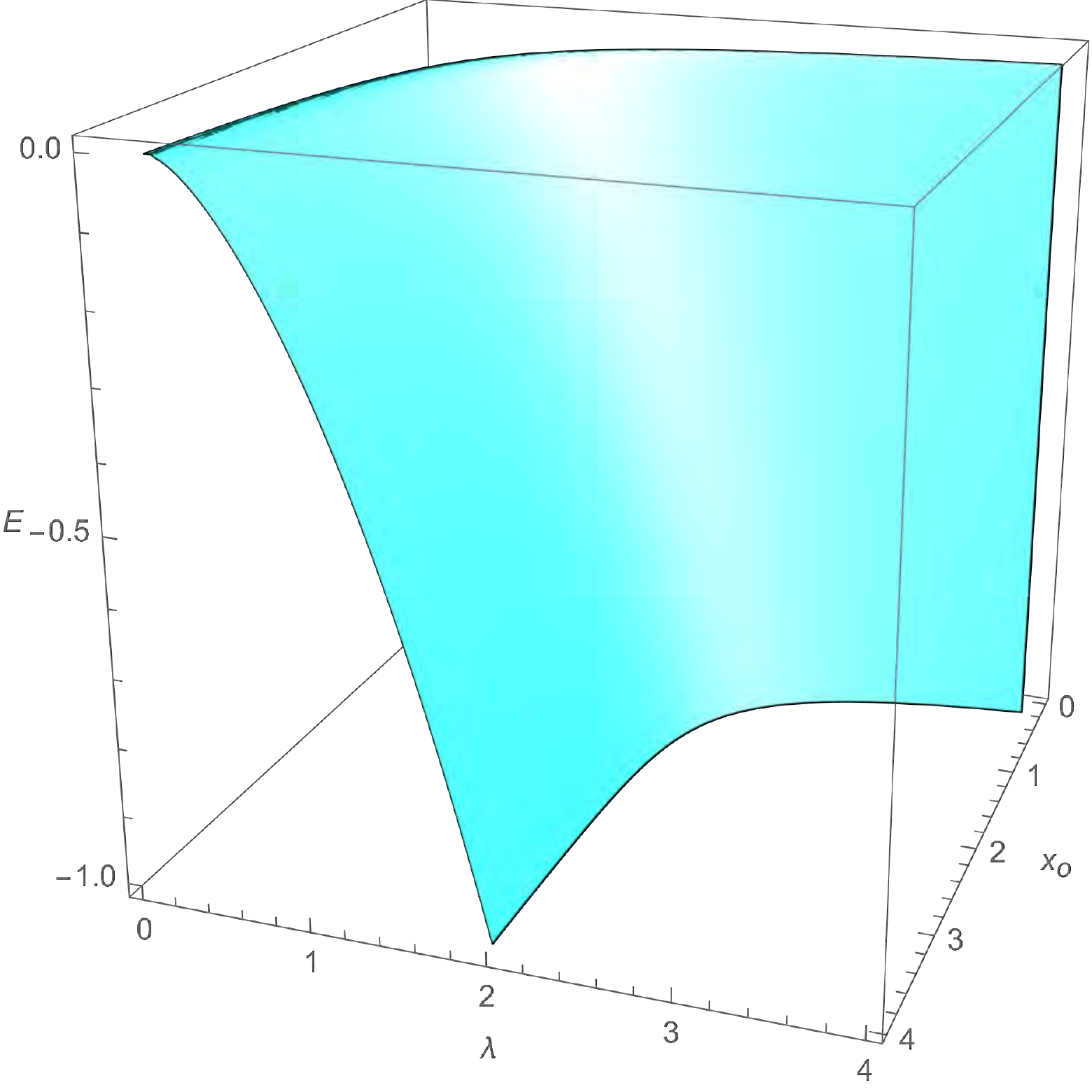}
\caption{The  energy $E_{D}(\lambda,x_0)$ as a function of $\lambda$ and $x_0$ given implicitly in \eqref{2.10}.}
\label{Dirichlet3D}
\end{center}
\end{figure}

In addition, we may keep $x_0$ fixed in \eqref{2.10} in order to get the eigenenergy, $E_{D}(\lambda)$, as a function of $\lambda$. Then, we obtain the plot shown on the right hand side of Figure~\ref{Dirichlet2D} for some particular values of $x_0$. Taking into account that the Hamiltonian $H^D_{\lambda,x_0}$ approaches $\left[-\frac {d^2}{dx^2}\right]_{D^+}$ as $\lambda\rightarrow 0$, it is almost obvious that $E_{D}(\lambda)\rightarrow 0$ as $\lambda\rightarrow 0$. Moreover after \eqref{2.10}, it is also clear that for large values of the coupling constant, $|E|^{1/2}$ must be large as well, so that $E_{D}(\lambda)\rightarrow -\infty$ as $\lambda\rightarrow +\infty$. Finally, we show  in Figure~\ref{Dirichlet3D} a three-dimensional plot of  the  energy $E_{D}(\lambda,x_0)$.

\subsection{Resonances}
\begin{figure}[hbtp]
\begin{center}
\includegraphics[width=0.34\textwidth]{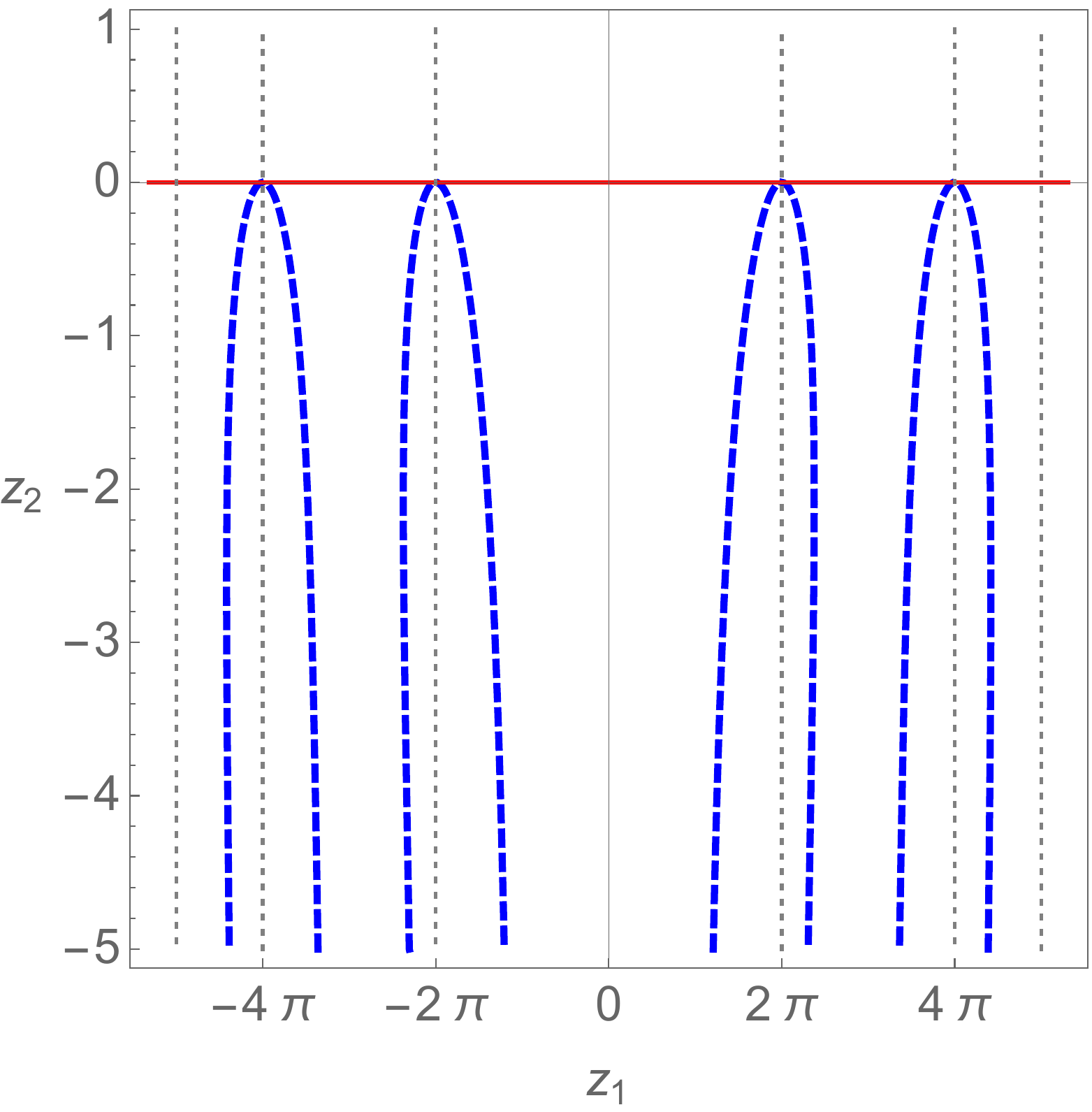}
\caption{The  resonances in the case of the Dirichlet problem, as given in \eqref{2.19}. Resonances lie over the dashed blue lines.}
\label{resonances_Dirichlet}
\end{center}
\end{figure}

Next, we investigate the possible presence of resonances in the sense given to this term in the Introduction. Thus, resonances are given by poles of the analytic continuation of \eqref{2.10}. Let us rewrite \eqref{2.10} for positive energies and, then, let us choose the appropriate branch for the square root. This gives:
\begin{equation}\label{2.15}
\lambda \, \frac{1-e^{- 2x_0 \sqrt{-E}}}{2\sqrt{-E}}=\lambda \, \frac{1-e^{2i x_0 \sqrt{E}}}{-2i\sqrt{E}}=1\,.
\end{equation}
As for the analytic continuation the variable $E$ is complex, let us write the square root as $\sqrt{E}=k_1+ik_2$. Then, \eqref{2.15} takes the form:
\begin{equation}\label{2.16}
\lambda (1-e^{2x_0 (-k_2+i\, k_1)})= 2k_2-2i\, k_1, \qquad k_1,k_2\in \mathbb{R}\,.
\end{equation}
From \eqref{2.16}, we obtain the following system of two real valued equations:
\begin{equation}\label{2.17}
\lambda (1-e^{- 2x_0 k_2} \cos(2x_0 k_1))=2k_2\,,
\qquad
\lambda e^{- 2x_0 k_2} \sin(2x_0 k_1)=2k_1\,.
\end{equation}
Next, let us define the following {\it real} variables:
\begin{equation}\label{2.18}
z_1=2 x_0 k_1, \quad z_2=2 x_0 k_2, \quad \alpha= x_0 \lambda\,,
\end{equation}
so that \eqref{2.17} becomes
\begin{equation}\label{2.19}
\alpha=\frac{z_2}{1-e^{- z_2} \cos z_1}=\frac{z_1}{e^{- z_2} \sin z_1}\,,
\end{equation}
which contains a transcendental equation relating the variables $z_1$ and $z_2$ providing the position of resonances in the complex $k$-plane, in other words, in the momentum representation. Each of the resonances is characterised by a solution of \eqref{2.19}. These solutions come into pairs of the form $\pm z_1-iz_2$ with $z_2>0$. Each of these pairs accounts for a resonance. Since the poles of an analytic complex function are isolated points, the  number of resonances is countably infinite. This situation is absolutely compatible with one of the characterisations of resonances, precisely the most popular among physicists, viewing resonances as poles of the analytic continuation of the $S$-matrix in the momentum representation, $S(k)$. The described arrangement of the resonance poles is in agreement with the standard choice of causality conditions for resonance scattering \cite{NUSS,BOHM}. Figure~\ref{resonances_Dirichlet} shows a graph where you can see the resonance poles in the momentum representation, located along the dashed blue lines. 

\section{The operator $\left[-\frac {d^2}{dx^2}\right]_{N^+}$ perturbed by an attractive Dirac delta}

Let us consider the self-adjoint Hamiltonian $\left[-\frac {d^2}{dx^2}\right]_{N^+}$, i.e., the one-dimensional Laplacian on the positive half-line with a Neumann boundary condition at the origin. Its Green function is for any $x,y>0$ and $E<0$ (see \cite{DER}):
\begin{equation}\label{3.1}
\left[\left[-\frac {d^2}{dx^2}\right]_{N^+}\!\!+ |E|\right]^{-1}\!\!(x,y)=\frac {e^{- |E|^{1/2} |x-y|}+e^{- |E|^{1/2}(x+y)}}{2 |E|^{1/2}}\,.
\end{equation}
The integral kernel of the semigroup is given by (see \cite{DER} as well):
\begin{equation}\label{3.2}
e^{-t\left[-\frac {d^2}{dx^2}\right]_{N^+}}(x,y)=\frac {e^{- \frac {(x-y)^2}{4t}}+e^{- \frac {(x+y)^2}{4t}}}{2 \sqrt{\pi t}}\,.
\end{equation}
As in the previous case, let us perturb $\left[-\frac {d^2}{dx^2}\right]_{N^+}$ with an attractive Dirac delta potential centred at $x_0$ with coupling constant $\lambda>0$, so that the new Hamiltonian becomes:
\begin{equation}\label{3.3}
H_{\lambda,x_0}^{N}=\left[-\frac {d^2}{dx^2}\right]_{N^+}\!\!-\lambda \delta (x-x_0)\,.
\end{equation}
By means of the same procedure used in the previous Section, we can obtain the explicit expression for the integral kernel of the resolvent of $H_{\lambda,x_0}^{N}$, which for $E<0$ is given by
\begin{eqnarray}\label{3.4}
(H_{\lambda,x_0}^{N}-E)^{-1}(x,y)= \left[\left[-\frac {d^2}{dx^2}\right]_{N^+}\!\!-\lambda \delta (x-x_0)+ |E|\right]^{-1}\!\! (x,y)  = \left[\left[-\frac {d^2}{dx^2}\right]_{N^+}\!\!+ |E|\right]^{-1}\!\!(x,y) \qquad
\nonumber\\[2ex]
+ \frac { \lambda}{1-\lambda \left[\left[-\frac {d^2}{dx^2}\right]_{N^+}\!\!+ |E|\right]^{-1}\!\! (x_0,x_0)}   \left[\left[-\frac {d^2}{dx^2}\right]_{N^+}\!\!+ |E|\right]^{-1}\!\! (x,x_0)  \left[\left[-\frac {d^2}{dx^2}\right]_{N^+}\!\!+ |E|\right]^{-1} \!\!(x_0,y)\,.
\end{eqnarray}
This result given by \eqref{3.4} coincides with the result obtained by using the Krein formula, see \cite{AGHH}. Then, since the eigenvalues of a self-adjoint operator are given by real poles of the resolvent, the equation determining these eigenvalues is 
\begin{equation}\label{3.5}
\lambda \left[\left[-\frac {d^2}{dx^2}\right]_{N^+}\!\!+ |E|\right]^{-1}\!\!(x_0,x_0)=\lambda\ \frac {1+e^{- 2|E|^{1/2}x_0}}{2 |E|^{1/2}}=1\,.
\end{equation}
It is remarkable that the above equation is exactly (2.11) in \cite{FR1}, which is the equation determining the energy of the ground state of the self-adjoint Hamiltonian $-\frac {d^2}{dx^2}-\lambda [\delta (x+x_0)+\delta (x-x_0)]$. This Hamiltonian is densely defined in $L^2(\mathbb R)$, where $\mathbb R$ stands for the {\it whole} real line. We want to obtain the eigenvalues of the energy as functions of $x_0$, for any fixed value of $\lambda>0$. Observe that \eqref{3.5} can be rewritten as:
\begin{equation}\label{3.6}
x_0^{N}(E)=-\frac 1{2 |E|^{1/2}} \text{ln} \left(\frac {2 |E|^{1/2}}{\lambda}-1\right)\,.
\end{equation}

For any fixed value of $\lambda$, the function $x_0^{N}(E)$ exists only for $-\lambda^2\leq E< -\frac {\lambda^2}{4}$. Then, whilst $x_0^{N}(E) \rightarrow +\infty$ as $E \rightarrow -\frac {\lambda^2}{4}$ from below, $x_0^{N}(E) \rightarrow 0$ as $E \rightarrow -\lambda^2$ from above. Hence, given that $x_0^{N}(E)$ is a strictly increasing function over its domain $\left[-\lambda^2,-\frac {\lambda^2}{4}\right)$, its inverse $E_{N}(x_0)$ is well defined over the semibounded interval $ \left[0, +\infty \right)$ with range $\left[-\lambda^2,-\frac {\lambda^2}{4}\right)$. Therefore, assuming $\lambda$ to be fixed, the eigenvalue $E_{N}(x_0)< -\frac {\lambda^2}{4}$ exists for $ x_0\geq 0$ and $E_{N}(0)=-\lambda^2$.

The fact that $E_{N}(x_0) \rightarrow -\lambda^2_{+}$ as $x_0 \rightarrow 0_{+}$, can be further understood by considering the  limit as $x_0 \rightarrow 0_{+}$ of the integral kernel of the following rank one operator on $L^2 [0,+\infty)$:
\begin{eqnarray}\label{3.7}
\!\!\!\!\!\!\!\! && \frac { \lambda \left(\left[-\frac {d^2}{dx^2}\right]_{N^+}\!\!+ |E|\right)^{-1}\!\!(x,x_0)  \left(\left[-\frac {d^2}{dx^2}\right]_{N^+}\!\!+ |E|\right)^{-1}\!\!(x_0,y) }{1-\lambda \left(\left[-\frac {d^2}{dx^2}\right]_{N^+}\!\!+ |E|\right)^{-1}\!\!(x_0,x_0)} \nonumber
\\[2ex] 
\!\!\!\!\!\!\!\! &&\qquad \longrightarrow \frac { \lambda  \left(\left[-\frac {d^2}{dx^2}\right]_{N^+}\!\!+ |E|\right)^{-1}\!\!(x,0)  \left(\left[-\frac {d^2}{dx^2}\right]_{N^+}\!\!+ |E|\right)^{-1}\!\!(0,y) }{1-\lambda \left(\left[-\frac {d^2}{dx^2}\right]_{N^+}\!\!+ |E|\right)^{-1}\!\!(0,0)} 
= \frac { 4\lambda  \frac {\displaystyle e^{-|E|^{1/2} x}}{\displaystyle 2 |E|^{1/2}}  \frac {\displaystyle e^{-|E|^{1/2} y }}{\displaystyle 2 |E|^{1/2}} }{\displaystyle 1-\frac {\lambda}{|E|^{1/2}}}\,.
\end{eqnarray}

This shows that
\begin{equation}\label{3.8}
\lim_{x_0 \to 0_+} \left[-\frac {d^2}{dx^2}\right]_{N^+}\!\!-\lambda \delta (x-x_0) = \left[-\frac {d^2}{dx^2}\right]_{N^+}\!\!-\lambda \delta (x)\,,
\end{equation}
in the norm resolvent convergence sense defined in \cite{RSI}. As in the Dirichlet case, described in the previous Section, the limit is analogous to those discussed in the aforementioned papers \cite{SF,ENZ,AFR4,AFR,FR1,AFR1,AFR2,FPR}. The Hamiltonian in the right hand side in \eqref{3.8} is exactly the $x$-space counterpart of the Aronszajn-Donoghue Hamiltonian of type II considered in 5.4.1 in \cite{DER}. Furthermore, as shown in the Dirichlet case, it is easy to prove that when the singularity placed at $x_0$ moves to infinity, the  value of the energy becomes $E_{N}(\lambda)=-\lambda^2/4$.

\begin{rem} We wish to warn the reader that, although the limit \eqref{3.8} is reminiscent of the one performed in \cite{FR1} proving that the negative one-dimensional Laplacian perturbed by two identical attractive deltas symmetrically placed about the origin converges to the Hamiltonian with a single attractive delta centred at the origin and having double strength, the operator obtained should not be regarded as the resolvent of the Hamiltonian with an attractive delta centred at the origin and having double strength. It is instead the resolvent of a self-adjoint Hamiltonian involving the combination of the Neumann condition and the $\delta$-condition at the origin. In fact, since the resolvent of the latter Hamiltonian is given by the sum of the resolvent of the negative Neumann Laplacian and the rank one operator \eqref{3.7}, it is clear that its domain consists of all the functions $\phi$ satisfying the Neumann condition $\phi'(0_{+})=0$ and those belonging to the one-dimensional subspace of the function $e^{-|E|^{1/2} x}$. For    $E=-\lambda^2$, the latter functions satisfy the boundary condition $\phi'(0_{+})=-\lambda \phi(0_{+})$.
\end{rem}

\begin{figure}[hbtp]
\begin{center}
\includegraphics[width=0.34\textwidth]{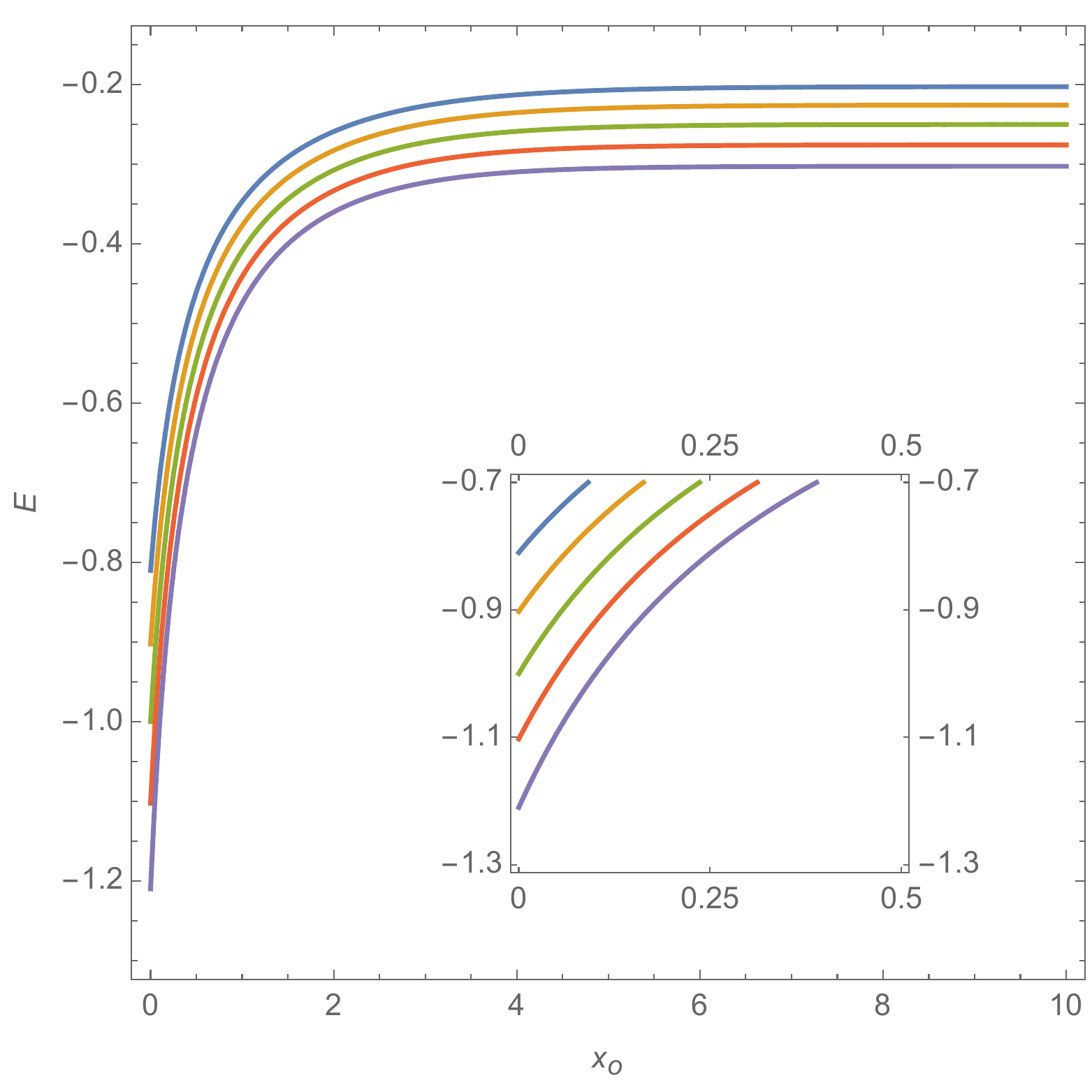}
\qquad\qquad\qquad
\includegraphics[width=0.34\textwidth]{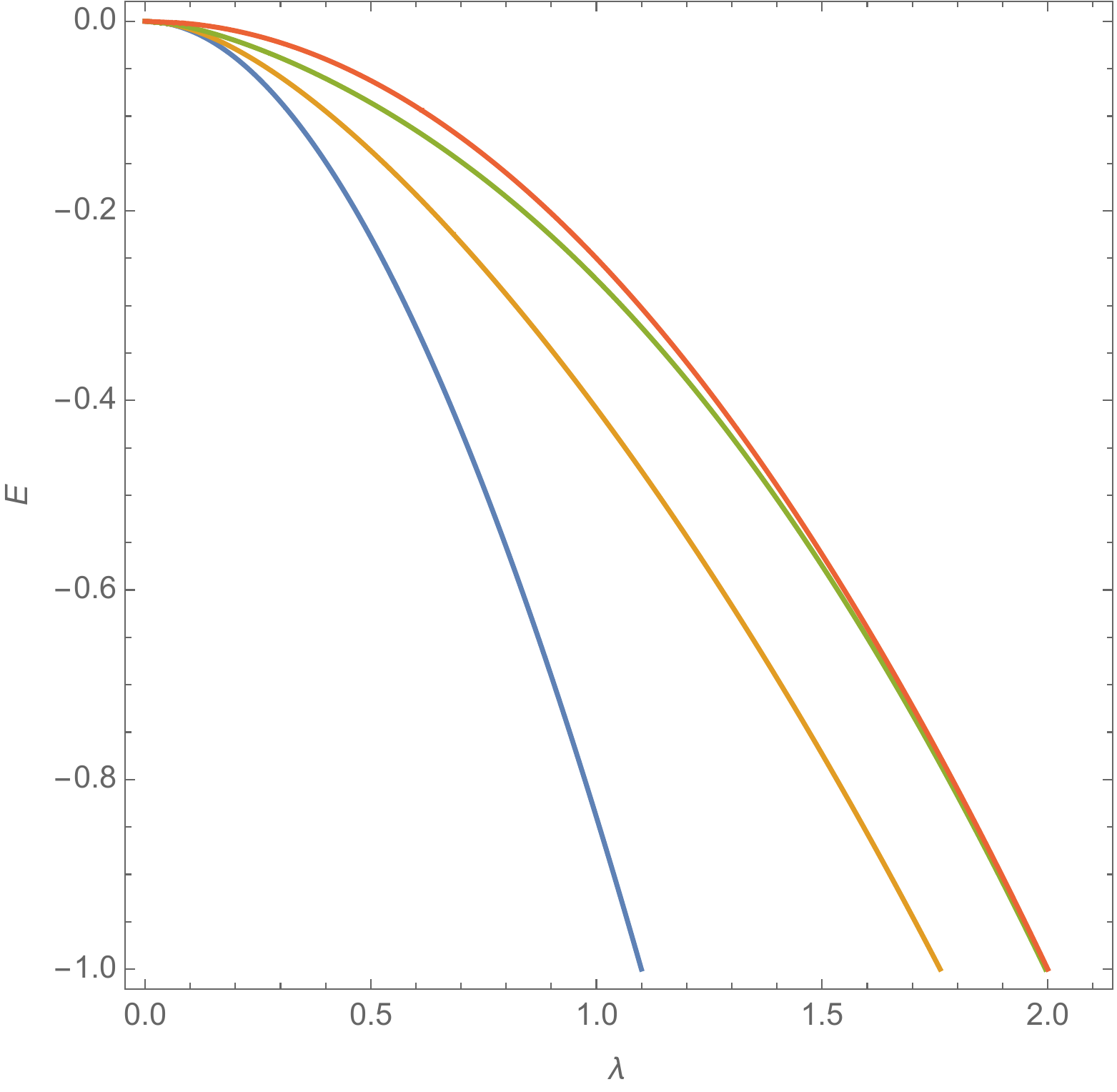}
\caption{ On the left,  the  energy $E_{N}(x_0)$ as a function of $x_0$ resulting from the solution of \eqref{3.5} for  
$\lambda=0.9$ (blue curve), $\lambda=0.95$ (orange curve), $\lambda=1$ (green curve), $\lambda=1.05$ (red curve), and $\lambda=1.1$ (violet curve). The inset shows the behavior of the curves near $x_0=0$.
On the right, the  energy $E_{N}(\lambda)$ as a function of $\lambda$ resulting from the solution of \eqref{3.5} for  
$x_0=0.1$ (blue curve), $x_0=1$ (orange curve), $x_0=3$ (green curve), and $x_0=\infty$ (red curve).}
\label{Neumann2D}
\end{center}
\end{figure}

\begin{figure}[hbtp]
\begin{center}
\includegraphics[width=0.34\textwidth]{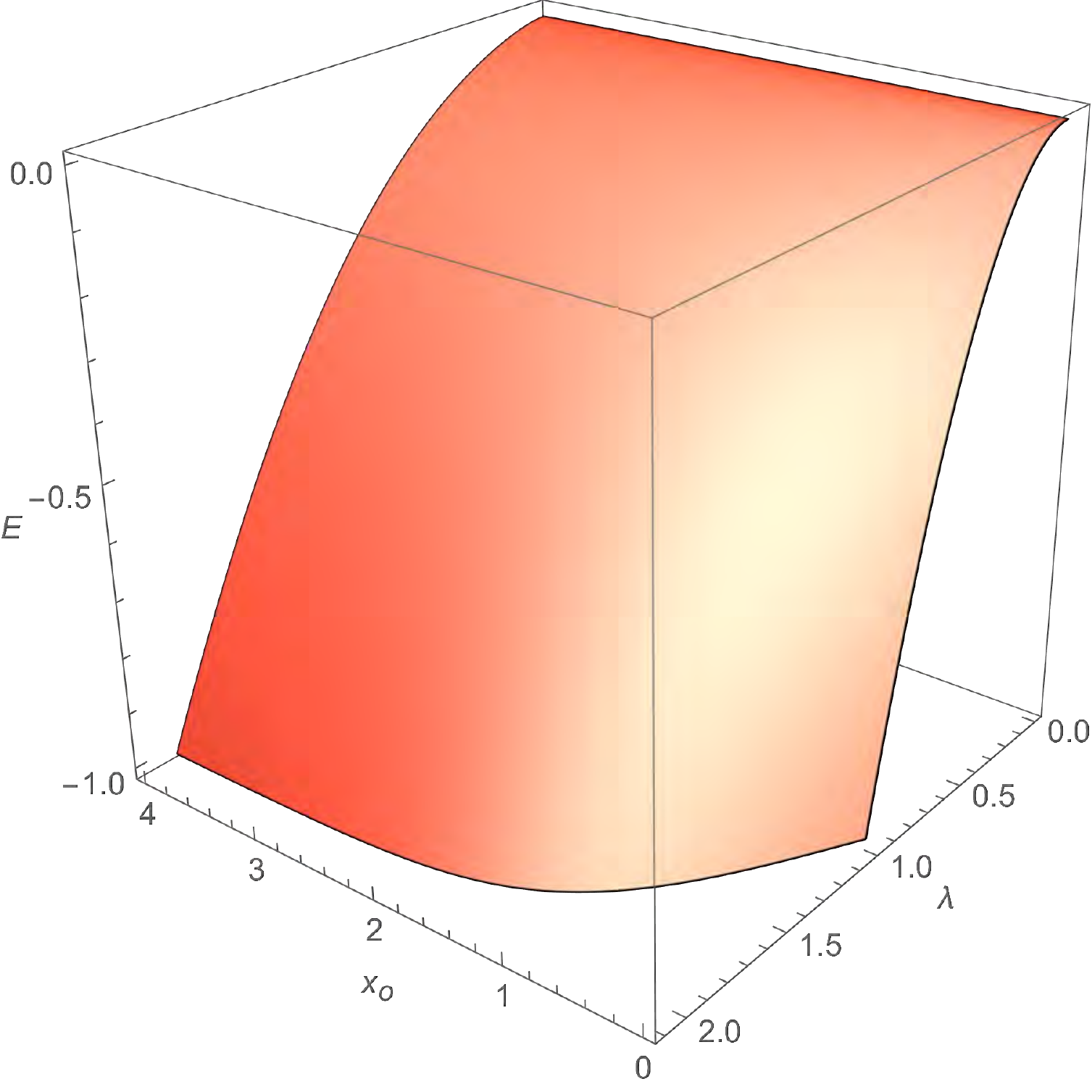}
\caption{The  energy $E_{N}(\lambda,x_0)$ as a function of $\lambda$ and $x_0$ as given implicitly by \eqref{3.5}.}
\label{Neumann3D}
\end{center}
\end{figure}

In addition, equation \eqref{3.5} can be studied by keeping $x_0$ fixed in order to get the energy value as a function of the coupling parameter $\lambda$, through the function $E_{N}(\lambda)$. Thus, we obtain the plot shown on the right hand side of Figure~\ref{Neumann2D} for some particular values of $x_0$. Taking into account that the Hamiltonian $H^N_{\lambda,x_0}$ approaches $\left[-\frac {d^2}{dx^2}\right]_{N^+}$ as $\lambda\rightarrow 0$, it is almost obvious that $E_{N}(\lambda)\rightarrow 0$ as $\lambda\rightarrow 0$. Moreover after \eqref{3.5}, it is also clear that for large values of $\lambda>0$, $|E|^{1/2}$ must be large as well, so that $E_{N}(\lambda)\rightarrow -\infty$ as $\lambda\rightarrow +\infty$. In Figure~\ref{Neumann3D}, we provide a complete three-dimensional plot of  the  energy $E_{N}(\lambda,x_0)$. 

\subsection{Resonances}
\begin{figure}[hbtp]
\begin{center}
\includegraphics[width=0.4\textwidth]{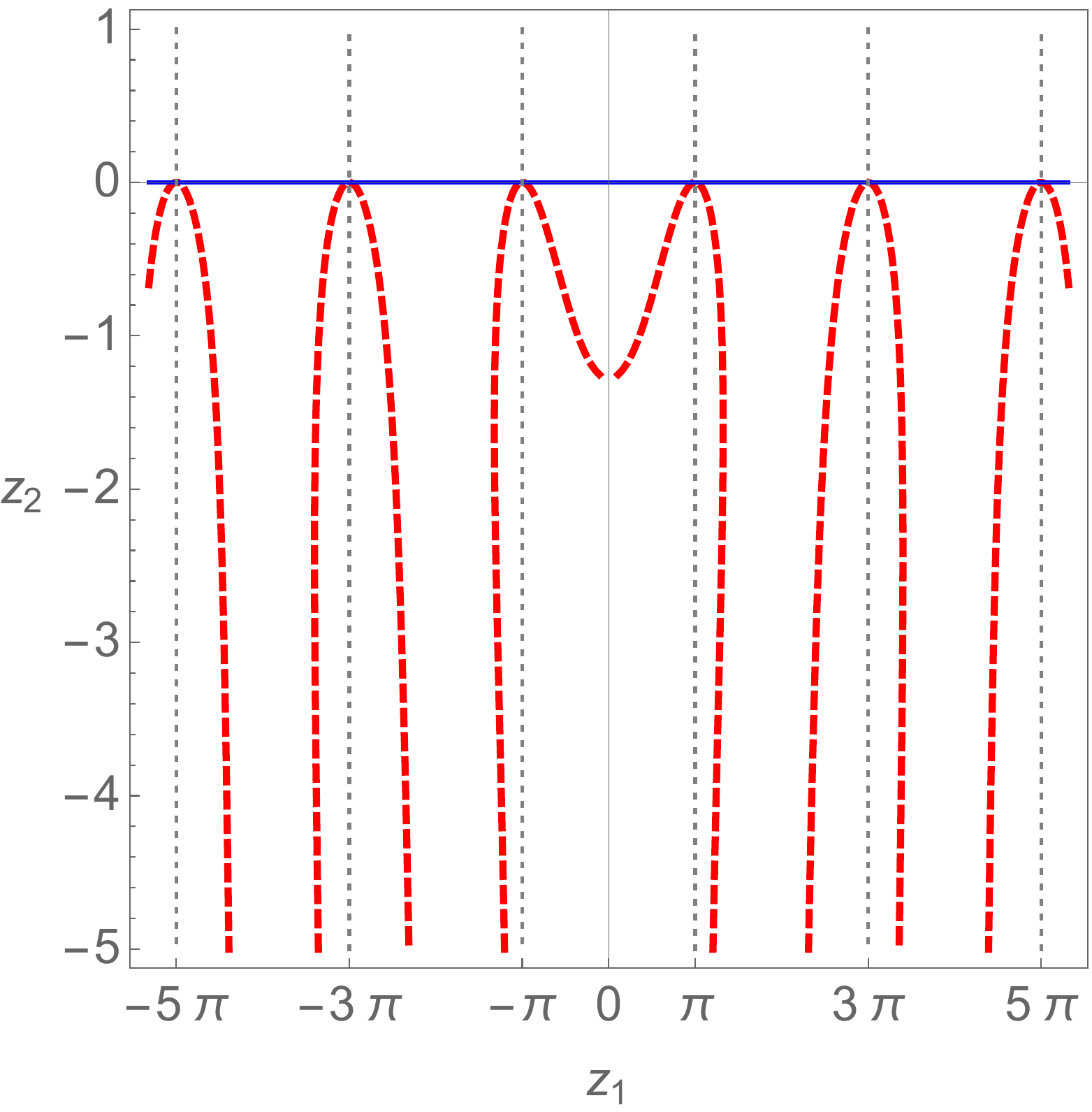}
\caption{The  resonances in the case of the Neumann problem, as given in \eqref{3.10}.}
\label{resonances_Neumann}
\end{center}
\end{figure}

We may also investigate the presence of resonances in this case. The procedure is exactly the same we have carried out for the Dirichlet case in the previous section. We have to find the poles of the analytic continuation of \eqref{3.5}, and also in the present case we shall obtain these poles in the so-called momentum representation. Let us rewrite \eqref{3.5} for positive energies and choose the appropriate determination of the square root as before. This yields the following relation:
\begin{equation}\label{3.9}
\lambda \, \frac{1+e^{- 2x_0 \sqrt{-E}}}{2\sqrt{-E}}=\lambda \, \frac{1+e^{2i x_0 \sqrt{E}}}{-2i\sqrt{E}}=1\,.
\end{equation}
As we have done in the previous section, let us take $E$ complex and write $\sqrt{E}=k_1+ik_2$, so that we arrive at the following relation:
\begin{equation}\label{3.10}
\alpha=\frac{z_2}{1+e^{- z_2} \cos z_1}=\frac{-z_1}{e^{- z_2} \sin z_1}\,,
\end{equation}
where $\alpha$ and $z_i$, $i=1,2$ have been already defined in \eqref{2.18}. This relation provides an infinite number of discrete solutions grouped into pairs of the form $\pm z_1-iz_2$ with $z_i>0$, $i=1,2$. These resonances lie on the dashed red lines in Figure~\ref{resonances_Neumann}.

\section{Application to the three-dimensional Laplacian perturbed by the attractive sphere interaction $- \lambda \delta (r-r_0)$}

As can be easily understood, differently from the three-dimensional Hamiltonian $-\Delta- \lambda \delta (\vec x\,)$ which requires either the theory of self-adjoint extensions or the renormalisation of the coupling constant in order to be rigorously defined \cite{AGHH,AK}, the Hamiltonian $-\Delta- \lambda \delta (r-r_0)$ can be defined in a far simpler way since it basically reduces to the study of a one-dimensional Laplacian operator with an attractive $\delta$-perturbation. The Hamiltonian $-\Delta- \lambda \delta (r-r_0)$ plays a crucial role in the so-called $\delta$-sphere interaction model, first analysed from a rigorous mathematical point of view in \cite{AGS} (see also \cite{DeLa} as well as the brief subsection on more general sphere interactions in \cite{AK}) despite having a rather long history in the more physically oriented literature, as pointed out in \cite{AGS}. We remind the reader that \cite{AGS} deals also with the $\delta'$-sphere interaction model. A recent detailed analysis of the model with the $a \delta (r-r_0)+b \delta' (r-r_0)$ sphere potential in $d$-dimensions ($d \geq 2$) can be found in \cite{CNR}. Detailed explanations of the crucial difference between the $\delta'$-interaction, that may not be added to the  $\delta$-interaction, and the $\delta' $-potential,  that instead may be added to the $\delta$-interaction, can be found in \cite{AFR3,AFR4,FGGN,LAN}. 

As a consequence of the possibility of reducing the study of $-\Delta- \lambda \delta (r-r_0)$  to a one-dimensional problem, the ground state energy of $-\Delta- \lambda \delta (r-r_0)=-\Delta_{\infty,0}- \lambda \delta (r-r_0)$ as well as that of $-\Delta_{0,0}- \lambda \delta (r-r_0)$, $-\Delta_{\infty,0}$ and $-\Delta_{0,0}$ being two remarkable self-adjoint extensions of the adjoint of the negative Laplacian initially defined on $C_0^{\infty}(R^3)$ (the rigorous definition of which can be found  in \cite{AGHH}), can be explicitly determined by exploiting the results of the previous sections. Before doing that, we wish to remind the reader that $-\Delta_{\infty,0}$ and $-\Delta_{0,0}$ are the very extensions used by Albeverio and collaborators in \cite{AHW} to investigate a class of exactly solvable problems in Quantum Mechanics and their relation to the Efimov effect. It is worth reminding the reader that the self-adjoint extension $-\Delta_{0,0}$ has the unique property of having a zero energy resonance, see \cite{AGHH, AFR2}. 

As follows from the thorough analysis carried out in \cite{AGHH}, the ground state of any self-adjoint extension of the negative Laplacian perturbed by $- \lambda \delta (r-r_0)$ is determined exclusively by the one-dimensional Hamiltonian $h_{0,\alpha}$ where
\begin{equation}\label{4.1}
h_{0,\alpha}\phi=-\frac {d^2 \phi}{dr^2}\,,
\end{equation}

for any absolutely continuous $\phi(r) \in L^2[0,\infty)$, such that $\frac {d \phi}{dr}$ is absolutely continuous, $\frac {d^2 \phi}{dr^2}\in L^2[0,\infty)$ and $\phi' (0_{+})=4 \pi \alpha \phi (0_{+})$. The operator $h_{0,\alpha}$ determines the properties of the three-dimensional Hamiltonian $-\Delta_{\alpha,0}$ on the subspace with zero angular momentum ($s$-wave). On the subspaces with angular momentum $ l\geq 1$, $-\Delta_{\alpha,0}$ takes the form
\begin{equation}\label{4.3}
h_{l}=-\frac {d^2}{dr^2}+ \frac {l(l+1)}{r^2}, \quad r>0, \ l=1,2, \dots\,.
\end{equation}
As pointed out in \cite{AGHH}, $h_{l},l\geq 1$ is self-adjoint on the subspace of $L^2[0,+ \infty)$ given by those absolutely continuous functions $\phi$ with absolutely continuous derivative such that $\phi' (0_{+})= \phi (0_{+})=0$ and
\begin{equation}\label{4.4}
-\frac {d^2 \phi}{dr^2}+ \frac {l(l+1)}{r^2}\phi
\end{equation}
is square integrable. For a detailed study of the Schr\"odinger equation with $\frac 1{r^2}$-potentials see \cite{NAR}. 

The above considerations imply that, the space of functions satisfying the above conditions with the boundary condition replaced by $\phi (0_{+})=0$ is the domain of self-adjointness domain for $h_{0,\infty}$. If we replace the boundary condition by $\phi' (0_{+})=0$, then, we obtain the domain of self-adjointness for $h_{0,0}$. 

As a consequence of the previous considerations, the ground state energy of 
$$
-\Delta- \lambda \delta (r-r_0)=-\Delta_{\infty,0}- \lambda \delta (r-r_0)
$$ 
coincides with $E_{D}(x_0)$, the single eigenvalue of  $\left[-\frac {d^2}{dx^2}\right]_{D^+} \!\! -\lambda \delta (r-r_0)$, which is also attested by the fact that the bound state equation \eqref{2.10} coincides with the one provided in  \cite{DeLa} and, in more general form, in \cite{CNR}, whereas the one of $-\Delta_{0,0}- \lambda \delta (r-r_0)$ is given by $E_{N}(x_0)$, the single eigenvalue of  $\left[-\frac {d^2}{dx^2}\right]_{N^+} \!\! - \lambda \delta (r-r_0)$.

Although the $\delta$-sphere model was not considered in \cite{AOV}, the authors of which chose a spherically symmetric Lorentzian potential peaked at $r=r_0$, it might be interesting to investigate whether a $\delta$-sphere interaction could be physically suitable to simulate a neutral fullerene such as $C_{60}$. Incidentally, the Lorentzian potential considered in \cite{AOV} might not be the most suitable choice since it does not belong to the Rollnik class (see \cite{RSII}) in three dimensions. If regarded as a one-dimensional potential on the positive half-line, the Lorentzian seems quite problematic since:
\begin{eqnarray}\label{4.5}
&& \left|\left| \frac 1{\sqrt{(r-r_0)^2+d^2}}\left[\left[-\frac {d^2}{dx^2}\right]_{D^+}\!\!+ |E|\right]^{-1}\frac 1{\sqrt{(r-r_0)^2+d^2}}\right|\right|_1= \nonumber\\[2ex] 
&& \qquad\qquad\qquad = \frac 1{2 |E|^{1/2}} \int_0^{\infty} \frac {1-e^{- 2|E|^{1/2}r}}{(r-r_0)^2+d^2} \, dr \rightarrow  \int_0^{\infty} \frac {r}{(r-r_0)^2+d^2} \, dr= \infty,
\end{eqnarray}
as $E \rightarrow 0$ from below. The latter result implies that the B-S operator is no longer trace class at $E=0$, so that the radial Schr\"odinger equation determining the ground state for $l=0$,
\begin{equation}\label{4.6}
-\frac {d^2f}{dr^2} -\frac {\lambda f}{(r-r_0)^2+d^2}= Ef\,,\quad \lambda>0\,,
\end{equation}
would admit no (respectively infinitely many) eigenvalues if $\lambda \leq \frac 1{4}$ (respectively $\lambda> \frac 1{4}$), due to some well-known results on this potential \cite{KlausHPA}.

\section{Concluding remarks}

In the present paper, we have investigated the behaviour of the single eigenvalue of the operators $\left[-\frac {d^2}{dx^2}\right]_{D^+}$ and $\left[-\frac {d^2}{dx^2}\right]_{N^+}$, that is, the one-dimensional Laplacians on the positive half-line with the respective Dirichlet and Neumann conditions at the origin, perturbed by an attractive Dirac distribution centred at a point $x_0> 0$. 

As can be expected due to the strongly repulsive nature of the Dirichlet condition at the origin, the eigenvalue produced by the attractive Dirac distribution can only emerge if its centre is sufficiently far away from the origin and its strength sufficiently large to make the attractive influence of the point interaction prevail over the strong repulsion exerted by the Dirichlet condition. The bound state equation we have determined makes this intuitive forecast quantitatively accurate: the eigenvalue arising from the attractive point interaction emerges provided $ \lambda x_0>1$, $ \lambda>0$ being the absolute value of the coefficient of the attractive Dirac distribution and $x_0>0$ its location along the positive semiaxis. It is rather remarkable that this bound state equation coincides with that determining the energy of the excited state of the self-adjoint Hamiltonian $-\frac {d^2}{dx^2}-\lambda [\delta (x+x_0)+\delta (x-x_0)]$, defined on a dense subspace the space of square integrable functions \textit{on the entire real line}, $L^2(\mathbb R)$. For any fixed value of $\lambda$, the eigenvalue $E_{D}(x_0)$ has been shown to emerge from the absolutely continuous spectrum at $x_0=\frac 1{\lambda}$ and decrease strictly to approach asymptotically the value $-\frac {\lambda^2}{4}$ from above.

As the Neumann condition is expected to be less repulsive than the Dirichlet one, it is not surprising that the eigenvalue produced by the attractive Dirac distribution in this case exists for any value of both parameters and is far more negative than the one arising in the Dirichlet case. It is rather noteworthy that the equation determining this eigenvalue coincides with that determining the energy of the ground state of the self-adjoint Hamiltonian $-\frac {d^2}{dx^2}-\lambda [\delta (x+x_0)+\delta (x-x_0)]$, defined on a dense subspace of the space of square integrable functions \textit{on the entire real line}, $L^2(\mathbb R)$.

In addition, we have also provided the eigenvalues for both Hamiltonians as functions of the magnitude of the coupling parameter for fixed $x_0$, the parameter specifying the location of the $\delta$-distribution, that is to say $E_{D}(\lambda)$ and $E_{N}(\lambda)$. Furthermore, both perturbed Hamiltonians show a countably infinite number of resonances defined as poles of the analytically continued resolvent of the total Hamiltonian, which are regular points of the resolvent of the unperturbed Hamiltonian.

The results regarding the above-mentioned one-dimensional Hamiltonians have been exploited in order to shed some light on the ground state energies of the three-dimensional operators  
$$
-\Delta- \lambda \delta (r-r_0)=-\Delta_{\infty,0}- \lambda \delta (r-r_0)\quad \text{and}\quad  -\Delta_{0,0}- \lambda \delta (r-r_0),
$$
 $-\Delta_{\infty,0}$ and $-\Delta_{0,0}$ being two remarkable self-adjoint operators studied in \cite{AGHH}.

\section*{Acknowledgements}

S. Fassari would like to thank Prof. Igor Yu. Popov and the entire staff at the Department of Mathematics, ITMO University, St. Petersburg for their warm hospitality throughout his stay. Stimulating discussions with A. Colli, M.Sc.(Damor Pharmaceuticals SpA, Naples, Italy) are kindly acknowledged by S. Fassari. S. Fassari's contribution to this work has been made possible by the financial support granted by the Government of the Russian Federation through the ITMO University Fellowship and Professorship Programme. This work was partially financially supported by the Government of the Russian Federation (grant 08-08) and by the Russian Science Foundation (grant 16-11-10330).
M. Gadella and  L.M. Nieto gratefully acknowledge partial financial support from Junta de Castilla y Le\'on and FEDER (Projects VA137G18 and BU229P18).

\end{document}